\documentclass[a4paper,reqno,12pt]{amsart}

\usepackage[dvips]{color}

\numberwithin{equation}{section}

\newcommand{\group}[1]{\mathsf{#1}}
\newcommand{\alg}[1]{\mathfrak{#1}}
\newcommand{\func}[2]{#1 \left( #2 \right)}
\newcommand{\brac}[1]{\left( #1 \right)}
\newcommand{\sqbrac}[1]{\left[ #1 \right]}
\newcommand{\set}[1]{\left\{ #1 \right\}}
\newcommand{\abs}[1]{\left| #1 \right|}
\newcommand{\ideal}[1]{\left< #1 \right>}
\newcommand{\inner}[2]{\left< #1 , #2 \right>}
\newcommand{\form}[2]{\Omega^{#1} \left( #2 \right)}
\newcommand{\homot}[2]{\pi_{#1} \left( #2 \right)}
\newcommand{\homol}[3]{\group{H}_{#1} \left( #2 ; #3 \right)}
\newcommand{\cohom}[3]{\group{H}^{#1} \left( #2 ; #3 \right)}
\newcommand{\ZZ}{\mathbb{Z}}
\newcommand{\QQ}{\mathbb{Q}}
\newcommand{\RR}{\mathbb{R}}

\newcommand{\dd}{\mathrm{d}}
\newcommand{\tang}[1]{T \left( #1 \right)}
\newcommand{\chern}[2]{\mathrm{c}_{#1} \left( #2 \right)}
\newcommand{\todd}[2]{\mathrm{Td}_{#1} \left( #2 \right)}
\newcommand{\euler}[1]{\mathrm{e} \left( #1 \right)}
\newcommand{\Cox}{\text{h}}
\newcommand{\dCox}{\Cox^{\vee}}
\newcommand{\tenslu}[3]{#1_{#2}^{\phantom{#2} #3} \,}

\DeclareMathOperator{\Hom}{Hom}
\DeclareMathOperator{\lcm}{lcm}

\newtheorem{theorem}{Theorem}

\begin{document}

\title[D-Brane Charges in WZW Models]{A Note on the Equality of 
Algebraic and\\  Geometric D-Brane Charges in WZW Models}

\author[P Bouwknegt]{Peter Bouwknegt}

\address[Peter Bouwknegt]{Department of Physics and 
Mathematical Physics, and Department of Pure Mathematics \\
University of Adelaide \\
Adelaide, SA 5005 \\
Australia}

\email{pbouwkne@physics.adelaide.edu.au, 
       pbouwkne@maths.adelaide.edu.au}

\author[D Ridout]{David Ridout}

\address[David Ridout]{Department of Physics and Mathematical Physics \\
University of Adelaide \\
Adelaide, SA 5005 \\
Australia}

\email{dridout@physics.adelaide.edu.au}

\thanks{PB is financially supported by the Australian Research Council,
and DR would like to thank La Trobe University for hospitality during 
various stages of this project.}

\begin{abstract}
The algebraic definition of charges for symmetry-preserving D-branes
in Wess-Zumino-Witten models is shown to coincide with the geometric
definition, for all simple Lie groups.  The charge group for such
branes is computed from the ambiguities inherent in the geometric
definition.
\end{abstract}

\maketitle

\section{Introduction} \label{secintro}

We consider symmetry-preserving D-branes in a Wess-Zumino-Witten model
on a compact, connected, simply-connected, simple Lie group $\group{G}$.  Let
$\group{T}$ be a maximal torus, and let $\alg{g}$ and $\alg{t}$ be the
respective Lie algebras.  Classically, the symmetry-preserving
D-branes of the WZW model on $\group{G}$ at level $k$ coincide, as
submanifolds, with certain conjugacy classes.  The quantised classes
are in bijection with the integrable highest weight modules at level
$k$ of the associated affine Lie algebra $\hat{\alg{g}}$.

Every conjugacy class intersects any chosen maximal torus in a finite
number of points.  If $h \in \group{T}$ is such a point for a
conjugacy class corresponding to a (symmetry-preserving) D-brane, then
we may identify the brane manifold with $\group{G} / \group{Z}$,
embedded in $\group{G}$, where $\group{Z}$ is the centraliser of $h$
\cite{AleDBr}.  When $h$ is a regular element, $\group{Z} =
\group{T}$, and we will call the brane regular.  These are the
symmetry-preserving branes of maximal dimension, and correspond to
weights (via the exponential map) in the interior of the affine
fundamental alcove\footnote{To be precise, we mean the simplex in
$\alg{t}^*$ corresponding to the affine fundamental alcove.} (at level
$k$).  When $h$ is singular, $\group{Z} \supset \group{T}$, and we
refer to these (lower dimensional) branes as singular.  They
correspond to weights on the boundary of the fundamental alcove.

There are at least two types of definition of D-brane charge in the
literature.  In \cite{FreBra}, Fredenhagen and Schomerus introduced
what we shall refer to as the algebraic definition, based on the
conformal field theory approach ``coupled'' with renormalisation flow
methods.  If $\hat{\lambda}$ is the highest weight of the integrable
highest weight module labelling a D-brane, then their definition
amounts to the charge
\begin{equation} \label{eqnalgcharge}
\func{Q_{\text{alg}}}{\lambda} = \dim \func{L}{\lambda},
\end{equation}
where $\func{L}{\lambda}$ is the irreducible highest weight module of $\alg{g}$
with (dominant integral) highest weight $\lambda$, the projection of
$\hat{\lambda}$ onto the weight space $\alg{t}^*$ of the horizontal
subalgebra.  To be precise, their charge is this dimension modulo some
integer which needs to be determined from the fusion rules.  This
integer was determined in \cite{FreBra,MalDBr} for symmetry-preserving
branes in (supersymmetric) WZW models on $\func{\group{SU}}{n}$, and
for the other simple Lie groups in \cite{BouDBr02}.  For instance,
such branes on $\func{\group{SU}}{2}$ admit charges valued modulo
$k+2$.  This charge definition appears to fit in nicely with the
notion that D-brane charges should be classified by twisted K-theory,
in that the charge groups are in agreement with the relevant
K-theoretic results which are presently known
\cite{MalDBr,BraTwi,FreTwi}.

By contrast, the geometric definition is a little more involved, and
we shall review its construction, and a little of its history, in
Section \ref{secgeocharge}.  We take some care with this construction, in
particular with the ``$\func{\group{U}}{1}$ flux'', so as to isolate the
ambiguities involved in defining the geometric brane charge.  In Section \ref{secSU3comp}, we compute
geometric charges for the regular branes in $\func{\group{SU}}{3}$ in
order to provide a non-trivial example (that is, not
$\func{\group{SU}}{2}$) showing that the geometric charge coincides
with the algebraic charge.  This obviously suggests that the algebraic
and geometric charges might coincide for all (simple) groups.  Section
\ref{secalg=geom} is devoted to proving this.  In Section
\ref{secambig}, we return to the ambiguities in the geometric charge
definition and investigate what this implies for D-brane charge groups.

\section{Geometric Charges} \label{secgeocharge}

WZW models come equipped with an integral $3$-form $H \in
\cohom{3}{\group{G}}{\ZZ} \cong \ZZ$ of period $k$, and each D-brane
(labelled by $\lambda$) comes with a $2$-form $\omega_{\lambda} \in
\form{2}{\group{G} / \group{Z}}$ whose derivative is $H$ restricted to
the D-brane \cite{KliOpe}.  Locally then $H = \dd B$, so $F_{\lambda}
= B - \omega_{\lambda}$ represents a degree-two cohomology class
wherever it is defined.  $F_{\lambda}$ is sometimes referred to as the $\func{\group{U}}{1}$ flux \cite{BacFlu}.  The coupling of the Ramond-Ramond fields in
string theory \cite{PolDir,BacFlu} then suggests a charge
\begin{equation} \label{eqngeocharge1}
\func{Q'_{\text{geo}}}{\lambda} = \int_{\group{G} / \group{Z}} e^{F_{\lambda}},
\end{equation}
where the integration is over the entire brane $\group{G} / \group{Z}$.
This definition has been known for some time to require modification,
and a careful analysis of certain quantum anomalies suggests the
correct modification \cite{MinKTh}.  We will consider
the modified charge shortly (Eqn. \eqref{eqngeocharge}), but first we
would like to examine the $2$-form $F_{\lambda}$ in more detail.

For the case $\group{G} = \func{\group{SU}}{2}$ whose D-brane
charges have been computed many times \cite{BacFlu,StaNot,AleRRC},
explicit calculations are elementary.  However, in general, there is
an issue regarding how one defines an appropriate $B$ on the entire
brane.  $H$ is exact there, but in the absence of further constraints
on $B$, adding a closed form to any candidate $B$ would allow us to
produce any numerical value we like for the geometric charge.

Let us instead follow \cite{StaNot,FigDBr} and note in that the
standard procedure leading to brane quantisation, integers associated
with each (allowed) brane naturally arise.  These are the possible
ambiguities in the action, and take the form
\begin{equation} \label{eqnactionambig}
\func{Q_{\text{amb}}}{\lambda , S^2 , M} = \int_M H - \int_{S^2} 
\omega_{\lambda},
\end{equation}
where $S^2$ is an arbitrary $2$-sphere in the brane and $M$ is an
arbitrary $3$-chain whose boundary is $S^2$.

If one considers $\func{\group{SU}}{2}$, whose group manifold is a
$3$-sphere, $S^2 = \func{\group{SU}}{2} / \func{\group{U}}{1}$ to be a
regular brane\footnote{The argument extends trivially to singular
branes by taking $S^2$ to be a degenerate $2$-sphere of zero radius.},
and $M$ to be a side of the $3$-sphere bounded by the brane,
$Q_{\text{amb}}$ can be easily computed and is found to agree
remarkably with the algebraic charge (to be precise, we find $Q_{alg} -
Q_{\text{amb}} = 1$).  Moreover, if we take $M$ to be the other side
of the brane, then $Q_{\text{amb}}$ changes by $k$, the period of $H$
over the $3$-sphere.  With the usual quantum shift, $k \rightarrow k +
\dCox$ ($\dCox$ the dual Coxeter number of $\alg{g}$), we conclude
that $Q_{\text{amb}}$ is therefore defined modulo $k+2$, again in
remarkable agreement with the algebraic result.  It appears then that
for $\func{\group{SU}}{2}$, $Q_{\text{amb}}$ defines a geometric
charge for branes.

Eqn. \eqref{eqngeocharge1} suggests that to generalise this to other
groups, we need to trivialise $H$, and exponentiate.  We shall do
exactly this to define $F_{\lambda}$, and thus
$\func{Q'_{\text{geo}}}{\lambda}$.  Consider therefore the brane
$\group{G} / \group{Z}$.  When $\group{Z}$ is the centraliser of a
torus\footnote{This is almost always true, as follows from a 
theorem of Steinberg \cite{SteEnd}.  In fact, at each level, there are 
at most $n + 1$ exceptions, where $n$ is the rank of $\group{G}$.  We are
chiefly concerned with regular branes, and so we shall not consider these
exceptions any further.},
the homology of $\group{G} / \group{Z}$ has no torsion and vanishes in
odd degrees \cite{BotApp}.  Choose a set $\set{S^2_i}$ of generators
of $\homol{2}{\group{G} / \group{Z}}{\ZZ}$ (we may take $2$-spheres as
generators by the Hurewicz isomorphism), and let $M_i$ be a $3$-chain
whose boundary is $S^2_i$ (since $\homol{2}{G}{\ZZ} = 0$, we may
choose them to be $3$-cells).  As the brane generally has
non-vanishing second homology group, the $M_i$ are not usually
contained within the brane.  Let $\mathcal{C}$ be the complex obtained
by attaching each $3$-cell $M_i$ to the brane along each corresponding
$S^2_i$.  Then $\homol{2}{\mathcal{C}}{\ZZ}$ clearly vanishes, and the
Mayer-Vietoris sequence for attaching \cite{BotDif} shows that
$\homol{3}{\mathcal{C}}{\ZZ}$ vanishes also.  As attaching $3$-cells
leaves the first homology group invariant, it follows from the
universal coefficient theorem that $\mathcal{C}$ has vanishing
cohomology in degrees $2$ and $3$.  Thus $B$ may be defined on
$\mathcal{C}$, so $F_{\lambda} = B - \omega_{\lambda}$ is defined on
the whole brane.  Note we are free to add any closed $2$-form on
$\mathcal{C}$ to $B$, but these are all exact on $\mathcal{C}$, hence
exact on the brane.  Thus $F_{\lambda}$ is well-defined in cohomology,
so $\func{Q'_{\text{geo}}}{\lambda}$ is also
well-defined\footnote{Given $\mathcal{C}$---see Section
\ref{secambig}.}.

Note that $\func{Q_{\text{amb}}}{\lambda , S^2_i , M_i}$ is just the
period of $F_{\lambda}$ over $S^2_i$.  A little Morse theory suggests
\cite{BotGeo} that for the $S^2_i$ we may take (translates of) certain
conjugacy classes of $\func{\group{SU}}{2}_{\alpha_i}$, the embedded
$\func{\group{SU}}{2}$-subgroups corresponding to the simple roots
$\alpha_i$.  The standard choice is now to take $M_i$ to be one side
of (the translated) $\func{\group{SU}}{2}_{\alpha_i}$, thus reducing
the computation of $Q_{\text{amb}}$ for general simple groups to the
$\func{\group{SU}}{2}$ case.  This is a standard computation and gives
$$\func{Q_{\text{amb}}}{\lambda , S^2_i , M_i} = 
\inner{\lambda}{\alpha_i^{\vee}}.$$
That is, the period of $F_{\lambda}$ over the 
chosen homology generator $S^2_i$ is just the Dynkin label $\lambda_i$.

As noted above, the algebraic and geometric charges are in close
agreement for $\func{\group{SU}}{2}$.  The computation for
$\func{\group{SU}}{3}$ does not seem to appear in the literature.  It
may be performed without fuss (for regular branes) using Schubert
calculus, and we shall outline it in Section \ref{secSU3comp}.  We
find that this time, the algebraic and (na\"{\i}ve) geometric charges differ by a
non-constant amount, suggesting that the geometric charge needs modifying.

As noted above, such a modification was suggested in \cite{MinKTh},
based on the cancellation of quantum anomalies.  There it was also
shown that their modified charge has a natural K-theoretic
interpretation.  Their result, after specialising to the case we are
interested in, amounts to
\begin{equation} \label{eqngeocharge}
\func{Q_{\text{geo}}}{\lambda} = \int_{\group{G} / \group{Z}}
e^{F_{\lambda}} \, \todd{}{\tang{\group{G} / \group{Z}}},
\end{equation}
where $\tang{M}$ denotes the tangent bundle of the manifold $M$, 
and $\todd{}{E}$ denotes the Todd class of the vector bundle $E$.

In \cite{AleRRC} this charge is computed for $\func{\group{SU}}{2}$
and is found to exactly agree with the algebraic charge (no
discrepancy of $1$).  In Section \ref{secSU3comp} we will extend this
to regular branes in $\func{\group{SU}}{3}$, using Schubert calculus
(although we need the Chern classes of the tangent bundle of the brane
manifold), and find again exact agreement.  We therefore adopt this
definition as our geometric brane charge.  In Section
\ref{secalg=geom}, we will prove that the algebraic and geometric
charges coincide for all simple Lie groups (but without invoking any
Schubert theory).

\section{$\func{\group{SU}}{3}$ Calculations} \label{secSU3comp}

We now digress to perform a couple of computations regarding regular
brane charges in $\func{\group{SU}}{3}$ using the technology of
Schubert calculus \cite{FulYou}.  This approach is convenient as the
brane manifolds are examples of (complete) flag manifolds.  The
independent homology cycles of the regular branes
$\func{\group{SU}}{3} / \func{\group{U}}{1}^2$, called Schubert cells,
are labelled by the elements of the Weyl group, $\group{W} =
\group{S}_3$, of $\func{\group{SU}}{3}$.  The $2$-spheres $S^2_i$
associated with the simple roots $\alpha_i$ (considered in Section
\ref{secgeocharge}), in fact correspond to the simple reflections
$w_i$.

The computational utility of Schubert calculus becomes apparent when
we take Poincar\'e duals and consider the cohomology ring.  The
Schubert cell $X_w \in \homol{\func{\ell}{w}}{\func{\group{SU}}{3} /
\func{\group{U}}{1}^2}{\ZZ}$ corresponding to $w \in \group{S}_3$
($\func{\ell}{w}$ is the length of $w$) has Poincar\'e dual denoted by
$p_{w_0 w} \in \cohom{\func{\ell}{w_0 w}}{\func{\group{SU}}{3} /
\func{\group{U}}{1}^2}{\ZZ}$, where $w_0 = w_1 w_2 w_1 = w_2 w_1 w_2$
is the longest element of $\group{S}_3$.  In turn, the $p_w$ may be
expressed as polynomials (called Schubert polynomials) in certain
generators $x_1 , x_2 \in \cohom{2}{\func{\group{SU}}{3} /
\func{\group{U}}{1}^2}{\ZZ}$.  These generators are chosen to be
naturally permuted by $\group{S}_3$.  To be specific, we have in the
usual representation,
\begin{align*}
p_e &= 1 & p_{w_1} &= x_1 & p_{w_2} &= x_1 + x_2 \\
p_{w_1 w_2} &= x_1 x_2 & p_{w_2 w_1} &= x_1^2 & 
p_{w_1 w_2 w_1} &= x_1^2 x_2.
\end{align*}
Finally, we also need the cohomology ring of $\func{\group{SU}}{3} /
\func{\group{U}}{1}^2$ which is a special case of a famous result of
Borel:
$$\cohom{*}{\func{\group{SU}}{3} / \func{\group{U}}{1}^2}{\ZZ} \cong
\frac{\ZZ \sqbrac{x_1 , x_2}}{\ideal{x_1^2 + x_1 x_2 + x_2^2 ,
x_1^3}}_.$$ The general result (Theorem \ref{thmBorel}) will be
important in Section \ref{secalg=geom}.  Note that the $p_w$ given
above generate the integral cohomology of $\func{\group{SU}}{3} /
\func{\group{U}}{1}^2$.

We will compute both $\func{Q'_{\text{geo}}}{\lambda}$ and
$\func{Q_{\text{geo}}}{\lambda}$ for regular branes in
$\func{\group{SU}}{3}$.  First, we determine $F_{\lambda}$ in terms of
the Schubert basis.  We have
$$\lambda_1 = \int_{X_{w_1}} F_{\lambda} = \int_{\func{\group{SU}}{3}
/ \func{\group{U}}{1}^2} F_{\lambda} \wedge p_{w_1 w_2} =
\int_{\func{\group{SU}}{3} / \func{\group{U}}{1}^2} F_{\lambda} \wedge
x_1 x_2$$ and similarly, $\lambda_2 = \int_{\func{\group{SU}}{3} /
\func{\group{U}}{1}^2} F_{\lambda} \wedge x_1^2$.  A little algebra
now gives
$$F_{\lambda} = \brac{\lambda_1 + \lambda_2} x_1 + \lambda_2 x_2.$$
Hence, making much use of the form of the cohomology ring,
\begin{equation} \label{eqnSU3geochargewrong}
\func{Q'_{\text{geo}}}{\lambda} = \frac{1}{3!}
\int_{\func{\group{SU}}{3} / \func{\group{U}}{1}^2} F_{\lambda}^3 =
\frac{1}{2} \brac{\lambda_1^2 \lambda_2 + \lambda_1 \lambda_2^2},
\end{equation}
which is not equal to the dimension of the irreducible representation with
highest weight $\lambda$.

We therefore turn to the computation of
$\func{Q_{\text{geo}}}{\lambda}$.  For $E$ a complex vector bundle,
the homogeneous components, $\todd{n}{E}$, of the Todd class
$\todd{}{E}$ can be expressed as polynomials (the Todd polynomials) in
the Chern classes, $\chern{i}{E}$, of $E$.  The first few are
\cite{HirTop}:
\begin{align*}
\todd{0}{E} &= 1 & \todd{1}{E} &= \frac{1}{2} \chern{1}{E} \\
\todd{2}{E} &= \frac{1}{12} \sqbrac{\chern{1}{E}^2 + \chern{2}{E}} &
\todd{3}{E} &= \frac{1}{24} \chern{1}{E} \chern{2}{E}.
\end{align*}
The contributing part of the integrand of
$\func{Q_{\text{geo}}}{\lambda}$ is therefore
\begin{multline*}
\todd{3}{E} + \todd{2}{E} F_{\lambda} + \frac{1}{2} \todd{1}{E}
F_{\lambda}^2 + \frac{1}{3!} \todd{0}{E} F_{\lambda}^3 \\ =
\frac{1}{24} \chern{1}{E} \chern{2}{E} + \frac{1}{12}
\sqbrac{\chern{1}{E}^2 + \chern{2}{E}} F_{\lambda} + \frac{1}{4}
\chern{1}{E} F_{\lambda}^2 + \frac{1}{6} F_{\lambda}^3,
\end{multline*}
where $E = \tang{\func{\group{SU}}{3} / \func{\group{U}}{1}^2}$,
considered as a complex vector bundle.

Evidently we need the Chern classes of this tangent bundle.  These may
be computed using a result of Borel and Hirzebruch which we shall use
in Section \ref{secalg=geom} (Theorem \ref{thmBorelHirze}).  We leave
it as an easy exercise to show from this result that\footnote{Note
that $\chern{3}{\tang{\func{\group{SU}}{3} / \func{\group{U}}{1}^2}} =
6 x_1^2 x_2$ is elementary, but is not required for this calculation.}
\begin{align*}
\chern{1}{\tang{\func{\group{SU}}{3} / \func{\group{U}}{1}^2}} &= 4
x_1 + 2 x_2 \\ \text{and} \qquad \chern{2}{\tang{\func{\group{SU}}{3}
/ \func{\group{U}}{1}^2}} &= 6 x_1^2 + 6 x_1 x_2.
\end{align*}
It immediately follows that 
\begin{align*}
\frac{1}{24} \chern{1}{E} \chern{2}{E} &= x_1^2 x_2, \\ \frac{1}{12}
\sqbrac{\chern{1}{E}^2 + \chern{2}{E}} F_{\lambda} &= \frac{3}{2}
\brac{\lambda_1 + \lambda_2} x_1^2 x_2, \\ \frac{1}{4} \chern{1}{E}
F_{\lambda}^2 &= \frac{1}{2} \brac{\lambda_1^2 + 4 \lambda_1 \lambda_2
+ \lambda_2^2} x_1^2 x_2, \\ \intertext{and from
Eqn. \eqref{eqnSU3geochargewrong} that} \frac{1}{6} F_{\lambda}^3 &=
\frac{1}{2} \brac{\lambda_1^2 \lambda_2 + \lambda_1 \lambda_2^2} x_1^2
x_2.
\end{align*}
Thus, the geometric charge is
\begin{align*}
\func{Q_{\text{geo}}}{\lambda} &= \int_{\func{\group{SU}}{3} /
\func{\group{U}}{1}^2} \left[ 1 + \frac{3}{2} \brac{\lambda_1 +
\lambda_2} + \frac{1}{2} \brac{\lambda_1^2 + 4 \lambda_1 \lambda_2 +
\lambda_2^2} + \frac{1}{2} \brac{\lambda_1^2 \lambda_2 + \lambda_1
\lambda_2^2} \right] x_1^2 x_2 \\ &= \frac{1}{2} \brac{\lambda_1 + 1}
\brac{\lambda_2 + 1} \brac{\lambda_1 + \lambda_2 + 2} = \dim
\func{L}{\lambda}.
\end{align*}
So the algebraic and geometric charges agree in this case.

\section{Proof:  The Charges Coincide Generally} \label{secalg=geom}

We now prove that the charges $\func{Q_{\text{alg}}}{\lambda}$ and
$\func{Q_{\text{geo}}}{\lambda}$ in fact coincide for
symmetry-preserving WZW D-branes in compact, connected, simply-connected, simple Lie groups.
Actually, for the purposes of clarity, we will only prove this for
regular branes.  The computation itself is also essentially contained in \cite{BorCha2}, though there the motivation is of course purely mathematical..  The extension to singular branes is left as an exercise 
for the interested reader.

The proof is based on two results which were alluded to in Section
\ref{secSU3comp}, one due to Borel \cite{BorCoh}, and the other to
Borel and Hirzebruch \cite{BorCha1}.  Before presenting these results,
it is convenient to introduce a formalism in which brane cohomology is
interpreted in terms of Lie-theoretic data.  We have a natural
sequence of isomorphisms,
$$\homol{2}{\group{G} / \group{T}}{\ZZ} \cong \homot{1}{\group{T}}
\cong \ker \set{ \exp : \alg{t} \rightarrow \group{T}} =
\group{Q}^{\vee},$$ where the first is transgression composed with the
isomorphism between $\homol{1}{\group{T}}{\ZZ}$ and
$\homot{1}{\group{T}}$, or alternatively, Hurewicz composed with the
isomorphism from the homotopy long exact sequence, and
$\group{Q}^{\vee}$ is the coroot (integral) lattice of $\alg{g}$.  It
follows now that in a very natural sense,
$$\cohom{2}{\group{G} / \group{T}}{\ZZ} =
\func{\Hom}{\homol{2}{\group{G} / \group{T}}{\ZZ} , \ZZ} \cong
\brac{\group{Q}^{\vee}}^* = \group{P},$$ where $\group{P}$ is the
weight lattice of $\alg{g}$.

Thus we may say that the fundamental weights generate the second
cohomology group of the regular branes.  With this formalism, Borel's
famous result on the cohomology of $\group{G} / \group{T}$ is:

\begin{theorem}[\cite{BorCoh}] \label{thmBorel}
Let $\Lambda_i$ denote the fundamental weights of $\group{G}$ (simple
of rank $r$), and $\group{W}$ denote the Weyl group of $\group{G}$.
Then the rational cohomology ring of $\group{G} / \group{T}$ is
generated by the $\Lambda_i$ (and the unit) modulo the
$\group{W}$-invariant polynomials of positive degree.  Specifically,
$$\cohom{*}{\group{G} / \group{T} }{\QQ} = \frac{\QQ \sqbrac{\Lambda_1
, \ldots , \Lambda_r}}{\mathcal{I}_+}_,$$ where $\mathcal{I}_+$ is the
ideal generated by the $\group{W}$-invariants of positive degree.

When the cohomology of $\group{G}$ has no torsion, that is $\group{G}
= \func{\group{SU}}{r+1}$ or $\func{\group{Sp}}{2 r}$, then the result
also holds over the integers.
\end{theorem}

The other result concerns the characteristic classes of the tangent
bundle to $\group{G} / \group{T}$.  Via the splitting principle
\cite{BotDif}, this bundle (treated as a complex vector bundle) is
cohomologically equivalent to a direct sum of line bundles.  The Chern
classes of the bundle are then the elementary symmetric polynomials in
the first Chern classes of the associated line bundles.

\begin{theorem}[\cite{BorCha1}] \label{thmBorelHirze}
With the above formalism, the first Chern classes of the line bundles
associated to $\tang{\group{G} / \group{T}}$ under the splitting
principle, are the positive roots of $\group{G}$.  Thus if $\Delta_+$
denotes the set of positive roots, the total Chern class of
$\tang{\group{G} / \group{T}}$ is given by
$$\chern{}{\tang{\group{G} / \group{T}}} = \prod_{\alpha \in \Delta_+}
\brac{1 + \alpha}.$$
\end{theorem}

We can now show that the algebraic and geometric charges coincide.
First however, note the beautiful interpretation of the cohomology
class of our $2$-form $F_{\lambda}$ in this formalism.  As the periods
of this form are just the Dynkin labels and the fundamental weights
are generators of $\cohom{2}{\group{G} / \group{T}}{\ZZ}$, it follows
that in this formalism, $F_{\lambda}$ is identified with $\lambda$.

Consider the integrand of the geometric charge.  We have \cite{HirTop}
\begin{align*}
e^{\lambda} \, \todd{}{\tang{\group{G} / \group{T}}} &= e^{\lambda}
\prod_{\alpha \in \Delta_+} \frac{\alpha}{1 - e^{- \alpha}} =
e^{\lambda} \frac{\prod_{\alpha \in \Delta_+} e^{\alpha /
2}}{\prod_{\alpha \in \Delta_+} \brac{e^{\alpha / 2} - e^{- \alpha /
2}}} \prod_{\alpha \in \Delta_+} \alpha \\ &= \frac{e^{\lambda +
\rho}}{\prod_{\alpha \in \Delta_+} \brac{e^{\alpha / 2} - e^{- \alpha
/ 2}}} \prod_{\alpha \in \Delta_+} \alpha,
\end{align*}
where $\rho = \frac{1}{2} \sum_{\alpha \in \Delta_+} \alpha$ is the
Weyl vector of $\group{G}$.

Note that $\prod_{\alpha \in \Delta_+} \alpha$ is cohomologically
non-trivial, for it is the top Chern class:
\begin{align*}
\int_{\group{G} / \group{T}} \prod_{\alpha \in \Delta_+} \alpha &=
\int_{\group{G} / \group{T}} \chern{m}{\tang{\group{G} / \group{T}}} =
\int_{\group{G} / \group{T}} \euler{\tang{\group{G} / \group{T}}} \\
&= \func{\chi}{\group{G} / \group{T}} = \abs{\group{W}},
\end{align*}
the last equality from the fact that the odd Betti numbers vanish and
the homology classes are in correspondence with $\group{W}$
\cite{BotGeo}.  Here $m = \abs{\Delta_+}$ (so $\dim \group{G} = 2 m +
r$), $\euler{E}$ is the Euler class of $E$ and $\func{\chi}{M}$ is the
Euler characteristic of $M$.

Now the only component of the integrand which contributes is that in
$\cohom{2 m}{\group{G} / \group{T}}{\RR}$.  Since $\prod_{\alpha \in
\Delta_+} \alpha$ generates this cohomology group, we should only have
to consider the degree-zero term in the prefactor
$$\frac{e^{\lambda}}{\prod_{\alpha \in \Delta_+} \brac{1 - e^{-
\alpha}}}_.$$ 
This prefactor, as a function on $\alg{t}$, has a pole
of order $m$ at the origin, and so it is not straight-forward to
extract its degree-zero term.  However, we recognise this as the
character of the Verma module for $\alg{g}$ of highest weight
$\lambda$.  This is an infinite-dimensional module, hence the
non-analytic behaviour.  Finite-dimensional modules, on the other
hand, have characters which are analytic everywhere.

With this in mind, let us note that the product of the roots is
anti-invariant under $\group{W}$:
$$\func{w}{\prod_{\alpha \in \Delta_+} \alpha} =
\brac{-1}^{\func{\ell}{w}} \prod_{\alpha \in \Delta_+} \alpha = \det w
\prod_{\alpha \in \Delta_+} \alpha.$$ As $\group{G} / \group{T}$ is
$\group{W}$-invariant (up to a change in orientation), we have
\begin{align*}
\int_{\group{G} / \group{T}} e^{\lambda} \prod_{\alpha \in \Delta_+}
\frac{\alpha}{1 - e^{- \alpha}} &= \det w \int_{\group{G} / \group{T}}
\func{w}{e^{\lambda + \rho} \prod_{\alpha \in \Delta_+}
\frac{\alpha}{e^{\alpha / 2} - e^{- \alpha / 2}}} \\ &= \det w
\int_{\group{G} / \group{T}} e^{\func{w}{\lambda + \rho} - \rho}
\prod_{\alpha \in \Delta_+} \frac{\alpha}{1 - e^{- \alpha}}_.
\end{align*}
We may therefore write
\begin{align*}
\int_{\group{G} / \group{T}} e^{\lambda} \ \todd{}{\tang{\group{G} /
\group{T}}} &= \frac{1}{\abs{\group{W}}} \int_{\group{G} / \group{T}}
\sum_{w \in \group{W}} \det w \ e^{\func{w}{\lambda + \rho} - \rho}
\prod_{\alpha \in \Delta_+} \frac{\alpha}{1 - e^{- \alpha}} \\ &=
\frac{1}{\abs{\group{W}}} \int_{\group{G} / \group{T}} \chi_{\lambda}
\prod_{\alpha \in \Delta_+} \alpha,
\end{align*}
where we recognise Weyl's character formula for the character,
$\chi_{\lambda}$, of the irreducible module $\func{L}{\lambda}$ with
highest weight $\lambda$.  As we now have a finite-dimensional module,
extracting the degree-zero term is trivial.  Obviously,
\begin{align*}
\func{Q_{\text{geo}}}{\lambda} &= \frac{1}{\abs{\group{W}}}
\int_{\group{G} / \group{T}} \chi_{\lambda} \prod_{\alpha \in
\Delta_+} \alpha = \frac{\dim \func{L}{\lambda}}{\abs{\group{W}}}
\int_{\group{G} / \group{T}} \prod_{\alpha \in \Delta_+} \alpha \\ &=
\dim \func{L}{\lambda} = \func{Q_{\text{alg}}}{\lambda},
\end{align*}
which completes the proof.

It is interesting to observe that in the integrand,
$$e^{\lambda + \rho} \prod_{\alpha \in \Delta_+}
\frac{\alpha}{e^{\alpha/2} - e^{- \alpha/2}}_,$$
the product is $\group{W}$-invariant, and so by
Theorem \ref{thmBorel}, it is cohomologically equivalent to
its zero-degree term.
Therefore, we may replace the product by a constant,
which is easily seen
to be in fact $1$.  It follows that for our regular
branes, the geometric
charge in fact reduces to
$$\func{Q_{\text{geo}}}{\lambda} = \int_{\group{G} /
\group{T}} e^{\lambda
+ \rho}.$$
That is, the na\"{\i}ve
geometric charge is only altered by the quantum shift
$\lambda \rightarrow
\lambda + \rho$.  Therefore, we have the
satisfying conclusion that the modified geometric
charge of \cite{MinKTh}, in this special case, reduces to the ``na\"{\i}ve''
charge of
\cite{PolDir} \emph{after quantisation is taken into
account}.  One can of
course check that the computation in Section
\ref{secSU3comp} is
consistent with this observation.

\section{Charge Group Constraints} \label{secambig}

Let us briefly review the constraints on the charge group derived from
the algebraic theory.  These charges were considered with regard to
brane condensation processes in \cite{FreBra}.  There, a formal
analogy with the Kondo model of condensed matter physics was exploited
to suggest charge conservation conditions of the form
\begin{equation} \label{eqnfusconstraints}
\func{Q_{\text{alg}}}{\lambda} \func{Q_{\text{alg}}}{\mu} = \sum_{\nu}
\tenslu{\mathcal{N}}{\lambda \mu}{\nu} \func{Q_{\text{alg}}}{\nu},
\end{equation}
where $\tenslu{\mathcal{N}}{\lambda \mu}{\nu}$ denotes the fusion
coefficients of the WZW theory (we assume symmetry-preserving branes
for simplicity).  These conditions are not usually satisfied over the
integers, and instead are interpreted as holding in the charge group
of the symmetry-preserving branes, which is of the form $\ZZ_x$.
Under the assumption that these constraints are exhaustive, the
integer $x$ was thereby determined for $\func{\group{SU}}{n}$ in
\cite{FreBra,MalDBr}, and general simple Lie groups $\group{G}$ in
\cite{BouDBr02} as
\begin{equation} \label{eqnchargegroupx}
x = \frac{k + \dCox}{\gcd \set{k + \dCox , y}}_,
\end{equation}
where $y \in \ZZ$ depends only on $\group{G}$ and is explicitly given
in Table 1 of \cite{BouDBr02} (recall $k$ is the level and $\dCox$ the
dual Coxeter number of $\alg{g}$).  Subsequently \cite{BraTwi}, the
computation of the relevant K-theories was reduced to a point where it
could be directly compared with the formalism of \cite{BouDBr02},
showing why the two approaches should agree.  This relied heavily on
the theorem of Freed, Hopkins and Teleman \cite{FreVer,FreTwi} which
says that twisted equivariant K-theory is in fact the fusion algebra. 

Consider now the geometric definition of brane charge.  We have
labelled our brane by a weight $\lambda$ which lies in the affine
fundamental Weyl alcove, and shown that the charge is given by $\dim
\func{L}{\lambda}$.  This labelling arises \cite{FelGeo} from the fact
that symmetry-preserving branes correspond to conjugacy classes
through some $h \in \group{T}$, where (after quantisation) 
$h=\exp \set{2 \pi i y}$, and $y \in \alg{t}$ is given as the image
of $(\lambda+\rho)/(k+\dCox)$ 
under the canonical isomorphism $\alg{t} \cong \alg{t}^*$.  Insisting
that $\lambda$ belong to the affine fundamental alcove fixes it
uniquely, which is why we use it to label our brane, but it is clear
that there are choices being made here.  Indeed, a cursory examination
of Sections \ref{secgeocharge} and \ref{secalg=geom} should convince
the reader that there is no reason why $\lambda$ must be taken in the
affine fundamental alcove.

First, $h \in \group{T}$, and therefore $y\in\alg{t}$, 
is only determined up to the action of
$\group{W}$.  Hence
$\lambda \in \alg{t}^*$ is ambiguous up to the shifted action of $\group{W}$,
$w \cdot \lambda = \func{w}{\lambda + \rho} - \rho$.
It follows that for the geometric charge to be
well-defined, we must have\footnote{Of course we must extend $\dim$,
in the obvious way, as a polynomial form on $\alg{t}^*$.  The
expression $\func{L}{\lambda}$ is not to be taken literally if
$\lambda$ is outside the fundamental chamber.}
\begin{equation} \label{eqnambigW}
\dim \func{L}{\lambda} = \det w \ \dim
\func{L}{w\cdot \lambda}
\end{equation}
for each integral $\lambda$ in the
fundamental alcove, and each $w \in \group{W}$.  The $\det w$ factor
arises as $w$ may reverse the orientation of the brane manifold.
It is an easy exercise to prove that
Eqn. \eqref{eqnambigW} is always satisfied (over $\ZZ$).

Next, $\exp : \alg{t} \rightarrow \group{T}$ is not injective, hence
$h \in \group{T}$ only determines $y \in \alg{t}$ up to the integral
lattice, which is the coroot lattice $\group{Q}^{\vee}$ (since
$\group{G}$ is simply-connected).  Thus $\lambda$ is also ambiguous up
to translations by $(k+\dCox)$ times the coroot lattice.  Together with the
first ambiguity, we therefore find that $\lambda$ is only determined
up to the shifted action of the \emph{affine} Weyl group (at level
$k$), $\widehat{\group{W}}_k$.  Hence for the geometric charge to be
well-defined,
\begin{equation} \label{eqnambigWhat}
\dim \func{L}{\lambda} = \det \widehat{w} \, \dim \func{L}{\widehat{w}
\cdot \lambda},
\end{equation}
for all integral $\lambda$ in the fundamental alcove (and hence all
integral $\lambda$), and each $\widehat{w} \in \widehat{\group{W}}_k$.
These relations do not hold in general over $\ZZ$ and should be
interpreted as constraints on the charge group $\ZZ_x$.

Recall that the affine Weyl group is used to compute the fusion
coefficients via the Kac-Walton formula \cite{WalFus,WalAlg}
$$\tenslu{\mathcal{N}}{\lambda \mu}{\nu} = \sum_{\substack{\widehat{w}
\in \widehat{\group{W}}_k \\ \widehat{w} \cdot \nu \in \group{P}_+}}
\det \widehat{w} \, \tenslu{N}{\lambda \mu}{\widehat{w} \cdot \nu},$$
where $\tenslu{N}{\lambda \mu}{\nu}$ are the tensor-product
multiplicities ($\func{L}{\lambda} \otimes \func{L}{\mu} \cong
\oplus_{\nu} \tenslu{N}{\lambda \mu}{\nu} \func{L}{\nu}$) and
$\group{P}_+$ is the set of dominant integral weights.  It therefore
seems reasonable to compare the charge groups obtained from the
constraints \eqref{eqnambigWhat} with those of
\eqref{eqnfusconstraints}.

For $\group{G} = \func{\group{SU}}{2}$ at level $k$,
\eqref{eqnambigWhat} admits the charge group\footnote{We have again
accounted for the quantum shift.} $\ZZ_{2 (k + 2)}$, whereas
\eqref{eqnfusconstraints} gives $\ZZ_{k + 2}$.  Numerical computations
for the other groups (low ranks and levels) suggest that in fact the
constraints \eqref{eqnambigWhat} reproduce exactly the charge groups
computed in \cite{BouDBr02}, except for $\func{\group{Sp}}{2n}$, $n$ a power of $2$ (note that $\func{\group{SU}}{2} = \func{\group{Sp}}{2}$).  As one
would expect, the exceptions noted there are not
reproduced\footnote{In light of comments in \cite{BraTwi,GabCha}, we
would like to point out that at low level, the constraints
\eqref{eqnfusconstraints}, by themselves, give rise to parameters $x$
which may differ from \eqref{eqnchargegroupx}, and that these
exceptions were \emph{explicitly} noted in \cite{BouDBr02} to imply
charge groups that are absurd from a K-theoretic viewpoint.
Implicitly then, the constraints \eqref{eqnfusconstraints}, as interpreted there, are not
completely exhaustive at sufficiently low levels.}.  When these charge
groups disagree, \eqref{eqnambigWhat} gives $\ZZ_x$ where $x$ is twice
what Eqn. \eqref{eqnchargegroupx} predicts.  This appears to be due to
fixed points of the $\widehat{\group{W}}_k$-action.  For instance, a
weight $\lambda$ on the boundary of the (shifted) affine alcove is
fixed by some reflection $\widehat{w} \in \widehat{\group{W}}_k$.
This leads to a constraint
$$2 \dim \func{L}{\lambda} = 0\  \mod x\,,$$ 
allowing this dimension to be
non-zero in $\ZZ_x$ if $x$ is even, whereas in the fusion ring,
boundary weights vanish.

There is still another ambiguity we have not accounted for yet.  Recall
in Section \ref{secgeocharge} that we constructed a complex
$\mathcal{C}$ by attaching $3$-cells to $\group{G} / \group{T}$ (for regular branes) along
a basis of homology $2$-spheres.  The $3$-cells correspond to
``halves'' of translated $\func{\group{SU}}{2}$-subgroups of
$\group{G}$, and are essentially chosen so that the periods of
$F_{\lambda}$ could be easily computed.  However, we may attach any
$3$-chains, with the correct boundaries, that we like (in particular we
could choose either half of the $\func{\group{SU}}{2}$-subgroups).
Whilst this will change $\mathcal{C}$, the periods of $F_{\lambda}$
can only change by a multiple of $(k+\dCox)$.  As $F_{\lambda}$ is identified
with $\lambda$ in the formalism of section \ref{secalg=geom}, we
conclude that for the purposes of $\func{Q_{\text{geo}}}{\lambda}$,
$\lambda$ is ambiguous up to translations by $(k+\dCox)$ times the
\emph{weight} lattice $\group{P}$ (for $\group{Z} = \group{T}$;
generally, only translations by a sublattice).  That is, for the geometric charge
to be well-defined, we must have
\begin{equation} \label{eqnambigweight}
\dim \func{L}{\lambda} = \dim \func{L}{\lambda + (k+\dCox)\mu},
\end{equation}
for all $\lambda , \mu \in P$.  Note that as $\group{Q}^{\vee}
\subseteq P$, the constraints \eqref{eqnambigweight} are stronger than
\eqref{eqnambigWhat}.

For $\func{\group{SU}}{2}$, we recover from these stronger constraints
the correct charge group $\ZZ_{k+2}$.  In fact, this ambiguity is what
was used in \cite{AleRRC,FigDBr} to compute this charge group.
Numerical computations for the other groups indicate that these
constraints \eqref{eqnambigweight} also reproduce exactly the charge
groups of \cite{BouDBr02}, except for $\func{\group{Sp}}{2n}$, this
time for $n$ \emph{not} a power of $2$ (now $x$ may be half of what
\eqref{eqnchargegroupx} predicts).  These weight lattice charge groups
were also investigated in \cite{BouDBr02} in the context of symmetries
of the charges obtained from \eqref{eqnfusconstraints}.  Specifically,
it was observed there that these charges displayed symmetries
corresponding to weight lattice translations for all groups except
$\func{\group{Sp}}{2n}$, $n$ not a power of $2$, and moreover, by
imposing such symmetries on $\func{\group{Sp}}{2n}$, one finds an
aesthetically pleasing universal formula for the integer $y$ of
Eqn. \eqref{eqnchargegroupx}, namely
$$y = \lcm \set{1 , 2 , \ldots , \Cox - 1},$$ where $\Cox$ is the
Coxeter number of $\alg{g}$.  The constraints \eqref{eqnambigweight}
thus provide a theoretical justification for this observation.

\vspace{1cm}

\begin{description}
\item[Note Added] After this note was prepared, another independent K-theory
computation appeared \cite{DouTwi} which gives explicit results for the
(universal covers of the) classical groups and $\group{G}_2$.  The torsion parts of these results appear to agree with the results, \eqref{eqnchargegroupx}, obtained from the dynamical constraints.
\end{description}

\end{document}